# Localização em Ambientes Internos Utilizando Redes Wi-Fi

David Alan de Oliveira Ferreira, Celso Barbosa Carvalho e Edjair de Souza Mota

*Resumo*—Este artigo apresenta um método de localização para ambientes internos capaz de melhorar a precisão de localização que é comprometida pela instabilidade do RSSI das redes IEEE 802.11. O método emprega o algoritmo $k$-Nearest Neighbors ($k$NN) e a análise de quartis para representar leituras de RSSI provenientes de APs (*Access Points*) de referência. Verificou-se que a proposta apresenta erro de localização nulo ao utilizar quatro APs e 10 leituras de RSSI de cada AP com apenas 0,69 segundo para localizar. Estes resultados são contribuições importantes, o que assegura que o método é promissor para localizar objetos em ambientes internos.

*Palavras-Chave*—*Localização interna, kNN, Wi-Fi, Quartis*

*Abstract*—This paper presents a localization method for indoor environments capable of improving the location accuracy that is hampered by instability in RSSI of the IEEE 802.11 networks. The method employs the $k$-Nearest Neighbors ($k$NN) algorithm and quartiles analysis in the data representation. The proposal had null error with only four APs and 10 readings per sample of each AP with just 0.69 second to locate. These values are important contributions, confirming that the method is promising to locate objects in indoor environments.

## I. Introdução

Uma das áreas de pesquisa em IoT (*Internet of Things*) é a determinação da localização dos objetos inteligentes, uma vez que tal informação pode disponibilizar novos serviços cientes de contexto [1, 2]. Embora o GPS (*Global Positioning System*) forneça dados aceitáveis da localização em ambientes externos, possui precisão comprometida em ambientes internos [3].

Diversos métodos na literatura podem ser utilizados para localizar objetos em ambientes internos [4-9]. Muitas destas propostas baseiam-se em redes Wi-Fi (*Wireless Fidelity*), devido à grande presença dessas redes em locais públicos e privados, e utilizam o RSSI (*Received Signal Strength Indicator*) como parâmetro para determinar a localização [4].

Os sistemas tradicionais de localização têm como base técnicas geométricas como a trilateração [5] ou utilizam padrões das leituras de RSSI como o *fingerprint*, porém, problemas de interferência nos sinais e instabilidade das medições, causados principalmente por obstáculos, dificultam a localização [6, 7].

Técnicas de alta acurácia, permitem implementar sistemas de localização em serviços como assistência médica, rastreamento de bens, marketing, etc. Novas abordagens estão sendo propostas com a utilização de métodos baseados em algoritmos de aprendizado de máquina a fim de localizar pessoas e objetos móveis [8, 9]. Neste sentido, pesquisas existentes na literatura [8, 9] têm criado técnicas com vistas à construção de conjuntos de dados adequados, tais como normalizar e representar dados brutos, identificar e remover dados redundantes; e criar modelos matemáticos. Contudo, verificou-se que tais pesquisas têm desconsiderado aspectos de variação das leituras de RSSI, a exemplo do grau de espalhamento dos dados quando afetados por diferentes obstáculos, resultando em baixa acurácia.

Levando isto em consideração, o objetivo desta pesquisa foi desenvolver um método de localização para ambientes internos baseado em aprendizado de máquina, utilizando atributos com informações específicas que caracterizam o comportamento do RSSI, a fim de melhorar os resultados de localização.

Este artigo está dividido nas Seções II, III e IV que apresentam, respectivamente: a metodologia, os resultados e discussão e as conclusões.

## II. Metodologia

### A. Cenário de experimentação

O método de localização proposto neste artigo foi desenvolvido e testado em um ambiente de experimentação prático. Utilizou-se uma sala com 12,46 m², com quatro mesas encostadas no centro de cada parede e um armário encostado no canto de uma das paredes. Foram utilizados módulos NodeMCU [10] configurados no padrão 802.11g.

Conforme Figura 1, configurou-se 8 módulos NodeMCU no modo AP (*Access Point*), identificados como AP1 a AP8 e configurados no canal 4, o de menor ocupação no local. O nó a ser localizado foi configurado como STA e denominado WSTA (*Wireless Station*).

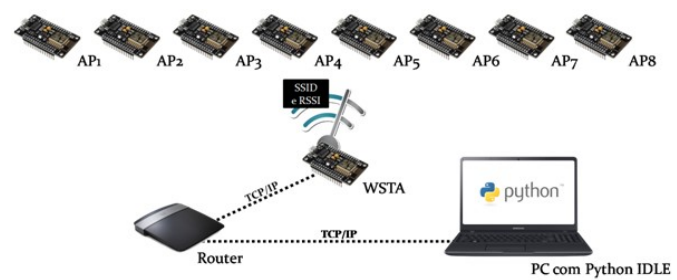

Fig. 1. Diagrama de conexões dos elementos que compõem o sistema de localização.

A WSTA varre os canais disponíveis a fim de identificar os APs existentes e coletar informações, como SSID (*Service Set IDentifier*) e o valor atual do RSSI, fundamentais para o sistema de localização. Configurou-se o PC com Python IDLE (*Python's Integrated Development and Learning Environment*) [11] para se conectar, via roteador sem fio (*Router*) a WSTA. No PC, desenvolveu-se programa para enviar requisições a WSTA, que possui servidor HTTP embarcado, a fim de obter informações de RSSI e SSID de cada um dos 8 APs do cenário

David Alan de Oliveira Ferreira e Celso Barbosa Carvalho, Faculdade de Tecnologia, Universidade Federal do Amazonas (UFAM), Manaus-AM, Brasil, Edjair de Souza Mota, Instituto de Computação, Universidade Federal do Amazonas (UFAM), Manaus-AM, Brasil, E-mails: ferreirad08@gmail.com, ccarvalho_@ufam.edu.br, edjair@icomp.ufam.edu.br.

de experimentação. Estas informações foram processadas por nosso algoritmo para obter a localização da WSTA.

Na Figura 2, visualizam-se detalhes do ambiente de experimentação com dimensões 3,50 m (eixo *X*), 3,56 m (eixo *Y*) e 2,80 m de altura (eixo *Z*). O cenário foi dividido em 16 zonas de mesmo tamanho e, no centro de cada zona (0,87 m de altura), posicionou-se RPs (*Reference Points*) para coleta de amostras de RSSI.

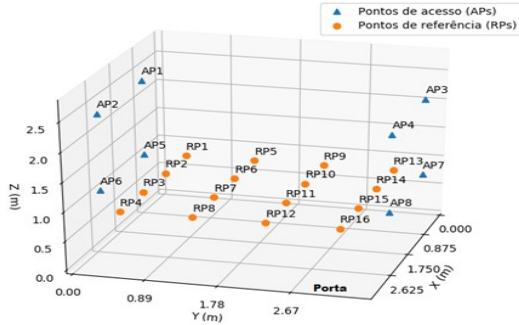

Fig. 2. Posicionamento dos pontos de acesso (APs) e dos pontos de referência (RPs) no ambiente interno.

### B. Experimentos para a relação entre distância e RSSI

Para verificar o comportamento da perda de potência dos sinais de rádio frequência (RF), inicialmente, utilizou-se a Equação 1 para determinar as distâncias entre cada AP e a WSTA posicionada sequencialmente nos RPs 1 e 2 (Figura 2).

$$d(p,q) = \sqrt{\sum_{i=1}^{3}(p_i - q_i)^2} \quad (1)$$

Onde: $p_i$ é o valor da coordenada *i* para o ponto *p* e $q_i$ é o valor da coordenada *i* para o ponto *q*.

Na Figura 3, testaram-se duas técnicas teóricas da literatura a fim de verificar a relação entre distância e RSSI: perda de propagação log-normal [12-14] e aproximação quadrática [15-18]. Na mesma figura, apresenta-se a média aritmética de 20 leituras de RSSI dos APs 1 a 8. Segundo [19], utilizar 20 leituras para determinar o RSSI médio é suficiente para reduzir o erro quadrático das variações.

Tanto para o RP1 (Figura 3a) quanto para o RP2 (Figura 3b), o RSSI decresceu exponencialmente com o aumento da distância nas técnicas perda de propagação log-normal e aproximação quadrática, conforme esperado. No entanto, o RSSI médio apresentou variações, não tendo uma relação estrita na proporção inversa do quadrado da distância. Assim, verificou-se que as técnicas teóricas tendem a ter resultados mais afastados do valor médio do RSSI neste cenário.

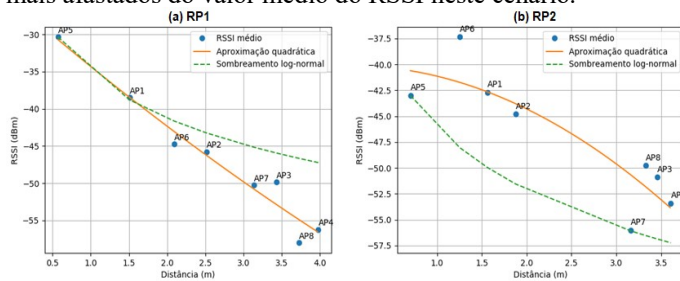

Fig. 3. Técnicas de perda de propagação log-normal, aproximação quadrática e RSSI médio para obtenção da intensidade do sinal recebido (RSSI).

As amostras de RSSI sofreram diferentes variações, decorrentes da ausência de linha de visada (*line of sight* – LOS) entre os APs e a WSTA (nó coletor) e dos fenômenos físicos de propagação das ondas eletromagnéticas causados por obstáculos presentes no ambiente. Portanto, as técnicas teóricas não expressaram as perdas de potência dos sinais em testes práticos.

Como o RSSI médio manteve este aspecto em testes conduzidos em dias diferentes, concluímos que os métodos relacionados com padrões disponibilizam mais informações sobre o RSSI. Desta forma, foram utilizados métodos baseados em *fingerprint* auxiliado por características específicas do comportamento do RSSI para determinar o RP em que a WSTA está posicionada.

### C. Propostas para as estimativas de localização

Visto que os RPs posicionados a diferentes distâncias de um AP podem apresentar o mesmo RSSI médio, é necessária a adição de características que os diferenciem. Uma alternativa é calcular os quartis [20] das leituras de RSSI, pois o segundo quartil (Q2 ou mediana) é mais robusto que a média aritmética quando há dados discrepantes (*outliers*) nas amostras [21].

Leituras de RSSI do AP2 coletadas nos RPs 1 e 2 (Figura 4a), distanciados 2,52 m e 1,88 m, respectivamente, resultaram no mesmo Q2 (-45 dBm) (Figura 4b), apesar das amostras terem sido coletadas em diferentes distâncias. No entanto, houve diferença no Q1, visto que menores valores de leituras de RSSI no RP1 estavam mais afastados do Q2 (Figura 4).

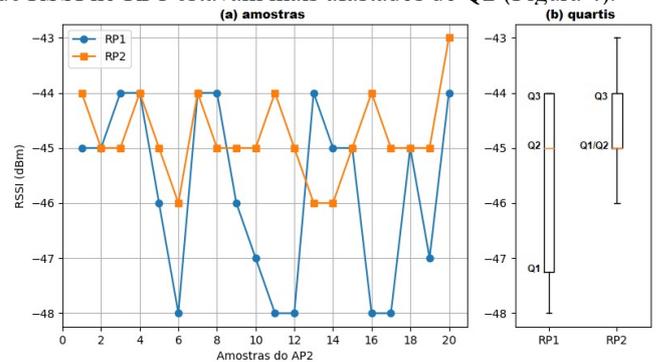

Fig. 4. Amostras da intensidade do sinal recebido (RSSI) e quartis do ponto de acesso 2 (AP2) nos pontos de referência 1 e 2 (RP1 e RP2).

O mesmo valor de Q2 de um AP em RPs com diferentes distâncias impossibilita o uso isolado dessa característica em algoritmos de classificação. Mas, a adição do primeiro e/ou o terceiro quartil (Q1 e Q3) tornam os dados adequados para algoritmos de classificação, como o algoritmo *k*NN [22].

*Localização utilizando as coordenadas do ponto de referência majoritário:*

Neste método de localização, o algoritmo de classificação *k*NN adota a distância euclidiana como função de similaridade e a análise de quartis na representação dos dados. Após transformação dos dados de RSSI em quartis e seleção dos *k* vizinhos mais próximos, estimou-se as coordenadas $(X_m, Y_m, Z_m)$ da WSTA pelas coordenadas do RP majoritário, ou seja, do RP com maior frequência de ocorrência dentre os *k* vizinhos mais próximos, conforme a Equação 2:

$$f_i = \max\{f_1, ..., f_n\} \rightarrow (X_m, Y_m, Z_m) = (X_i, Y_i, Z_i) \quad (2)$$

Isto é, se a frequência de ocorrência $f_i$ é a maior frequência observada, então $(X_m, Y_m, Z_m)$ são iguais as coordenadas $(X_i, Y_i, Z_i)$ do ponto de referência *i*. Havendo empate nas frequências de múltiplos RPs, atribui-se as coordenadas do RP associado ao vizinho mais próximo à localização estimada.

As coordenadas do RP majoritário são comumente utilizadas quando o algoritmo $k$NN é indicado para o problema de localização [23-25]. No entanto, a estimativa só é possível nas posições onde foram coletadas amostras de RSSI.

*Localização utilizando as coordenadas do centróide dos pontos de referência:*

Nesta abordagem, utilizou-se o algoritmo $k$NN adotando a distância euclidiana como função de similaridade e o conjunto de instâncias baseadas em quartis. Após transformação dos dados de RSSI em quartis e seleção dos $k$ vizinhos mais próximos, estimou-se a localização da WSTA nas coordenadas $(X_c, Y_c, Z_c)$. Sendo estas coordenadas calculadas através da média das coordenadas dos RPs $(X_i, Y_i, Z_i)$ associados aos $k$ vizinhos mais próximos, ponderadas pelas respectivas frequências das ocorrências de cada RP ($p_i$), conforme a Equação 3:

$$(X_c, Y_c, Z_c) = \left( \frac{\sum_{i=1}^{n} p_i X_i}{\sum_{i=1}^{n} p_i}, \frac{\sum_{i=1}^{n} p_i Y_i}{\sum_{i=1}^{n} p_i}, \frac{\sum_{i=1}^{n} p_i Z_i}{\sum_{i=1}^{n} p_i} \right) \quad (3)$$

Onde: $(X_c, Y_c, Z_c)$ são as coordenadas do centróide e $p_i$ é o peso para o ponto de referência $i$.

O centróide possibilita a estimativa de localização em qualquer posição do ambiente interno, incluindo posições onde não foram coletadas amostras de RSSI [12, 26].

### D. Experimentos para as estimativas de localização

Os métodos anteriores são chamados de Método I e Método II, respectivamente. A Figura 5 ilustra o diagrama de fluxo comum aos métodos propostos que envolvem duas fases.

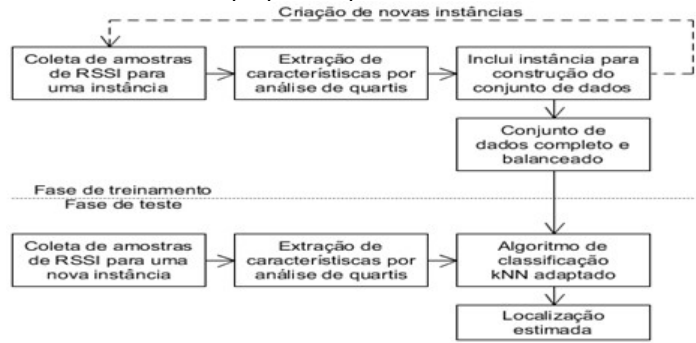

Fig. 5. Diagrama de fluxo comum aos métodos propostos.

*Fase de treinamento:*

Consiste na criação de um conjunto de instâncias previamente classificadas em múltiplas classes. As classes do conjunto de treinamento representam os 16 pontos de referência (RPs). Para criar uma instância previamente classificada em uma classe, a WSTA é posicionada manualmente no RP correspondente à classe, onde são coletadas amostras com $m=20$ leituras de RSSI bruto de cada um dos $n$ APs, e organizadas numa matriz $A_i$ ($m \times n$). Cada coluna contém $m$ leituras de determinado AP.

$$A_i = \begin{bmatrix} RSSI_{11} & \cdots & RSSI_{1n} \\ \vdots & \ddots & \vdots \\ RSSI_{m1} & \cdots & RSSI_{mn} \end{bmatrix} \quad (4)$$

Em seguida, os três quartis (Q1, Q2 e Q3) foram calculados para os $n$ APs com base nas suas respectivas amostras. O vetor $b_i$ representa uma instância formada pelos quartis calculados a partir das amostras (colunas) que compõem a matriz $A_i$, resultando em um vetor linha composto por $3n$ atributos.

$$b_i = [\overbrace{Q1 \quad Q2 \quad Q3}^{AP_1} \quad \overbrace{Q1 \quad Q2 \quad Q3}^{AP_2} \quad \cdots \quad \overbrace{Q1 \quad Q2 \quad Q3}^{AP_n}] \quad (5)$$

Para criar instâncias previamente classificadas em múltiplas classes, a WSTA foi posicionada sequencialmente em todos os 16 RPs e os passos supracitados foram repetidos. Com intuito de balancear o conjunto de treinamento, isto é, manter o mesmo número de instâncias entre as classes, criou-se 10 instâncias para cada uma das 16 classes.

No final desta etapa, obtive-se o conjunto de treinamento (matriz $T$ de dimensões $160 \times 24$) contendo 160 instâncias representadas por $b_i$ com 24 atributos, sendo três quartis das amostras de cada um dos oito APs, em que cada linha da matriz $T$ representa uma instância e cada coluna um atributo.

$$T = \begin{bmatrix} b_1 \\ \vdots \\ b_{10} \\ \vdots \\ b_{151} \\ \vdots \\ b_{160} \end{bmatrix} \begin{matrix} \} RP_1 \\ \vdots \\ \} RP_{16} \end{matrix} \quad (6)$$

*Fase de testes:*

Criou-se um conjunto de teste (matriz $E$) balanceado com 160 novas instâncias a partir de novas amostras de RSSI, coletadas nos mesmos 16 RPs. Os RPs do conjunto de teste são omitidos no processo de classificação e utilizados somente em análise posterior para verificar a precisão dos métodos.

$$E = \begin{bmatrix} b_1 \\ \vdots \\ b_{10} \\ \vdots \\ b_{151} \\ \vdots \\ b_{160} \end{bmatrix} \begin{matrix} \} RP_1 \\ \vdots \\ \} RP_{16} \end{matrix} \quad (7)$$

As instâncias da matriz $E$ foram aplicadas ao processo de classificação pelo algoritmo $k$NN adaptado aos métodos I e II, que possuía armazenado previamente o conjunto de treinamento $T$ (matriz 6). Os métodos de localização estimaram a posição real da WSTA associada a cada instância de teste e em seguida foi verificada a precisão das estimativas.

A fim de obter o menor erro de localização para cada método e com baixo custo computacional, o processo de classificação foi submetido a 49 tratamentos quantitativos, definidos por combinações, variando o número de APs ($n$) e o parâmetro $k$. O número de APs testados variou de 2 a 8, pois a utilização de apenas um AP resultaria em alta probabilidade de repetição de RSSI, acarretando em baixa precisão. O parâmetro $k$ foi testado com valores ímpares de 1 até 13.

Estes tratamentos permitiram mensurar o melhor número de APs e o número adequado de vizinhos mais próximos, ou seja, o melhor par ($n$, $k$). Quando o par ($n$, $k$) de melhor precisão é definido com os menores valores possíveis, tem-se um menor consumo de hardware e tempo de processamento reduzido, pois menos dados são coletados e tratados pelo algoritmo.

Foram realizadas 7840 *(160x49)* estimativas de localização para cada proposta (Métodos I e II) e feita comparação com os métodos 3-PCA [26] e *Powed*-Sørensen (PS) [23] realizados de forma prática na literatura. Devido as diferenças nos ambientes de testes utilizados pelos autores, quanto à dimensão do espaço, quantidade de RPs e mobília, estes métodos foram fisicamente testados no mesmo ambiente que os métodos propostos.

Por ambos estimarem a localização nas coordenadas do RP majoritário, o método I foi comparado ao método PS, que utiliza a função de similaridade Sørensen combinada com a representação de dados *Powed*. E por ambos estimarem a localização através do centróide dos RPs, o método II foi comparado ao método 3-PCA, que combina a distância euclidiana como função de similaridade com a PCA para representar o conjunto de dados (Tabela I).

TABELA I.   CARACTERÍSTICAS DOS MÉTODOS PROPOSTOS E DOS MÉTODOS APRESENTADOS NA LITERATURA

| | Método | Função de similaridade | Representação dos dados | Obtenção das coordenadas |
|---|---|---|---|---|
| Propostas | I | Distância euclidiana | Análise de quartis | RP majoritário |
| | II | Distância euclidiana | Análise de quartis | Centróide dos RPs |
| Literatura | 3-PCA | Distância euclidiana | PCA | Centróide dos RPs |
| | PS | Sørensen | *Powed* | RP majoritário |

## III. RESULTADOS E DISCUSSÃO

### A. Função de distribuição acumulada do erro médio de localização

Para analisar o desempenho dos quatro métodos a partir das estimativas de localização, independente do número de APs instalados ($n$) e do valor do parâmetro $k$, foi construído o gráfico da Função de Distribuição Acumulada (CDF) dos erros de localização. O erro médio (EM) foi calculado em metros, conforme a Equação 8:

$$EM = \frac{1}{N}\sum_{i=1}^{N} d(P_i, P_i^{'}) \qquad (8)$$

Onde: a distância entre $P_i$ e $P_i^{'}$ é o erro de cada localização, $P_i$ é a posição real da WSTA, $P_i^{'}$ é a posição estimada da WSTA e $N$ é o número de estimativas correspondentes.

O método I produziu os melhores resultados, onde todas as estimativas apresentaram EM inferior a 0,12 m, enquanto que os métodos II, 3-PCA e PS apresentaram EM menor que 0,12 m em apenas 73,47, 45,24 e 71,43% das estimativas, respectivamente. O método PS alcançou um EM máximo inferior aos erros dos métodos II e 3-PCA, porém o EM máximo do método PS (0,3233 m) é quase três vezes maior que o EM máximo do método I (0,1152 m) (Figura 6).

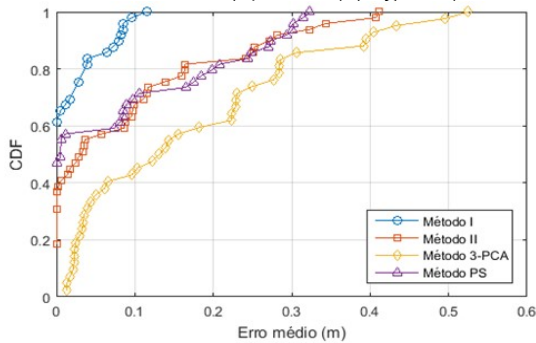

Fig. 6.   Função de Distribuição Acumulada dos erros de localização.

Os métodos II e 3-PCA apresentaram os piores desempenhos, possivelmente por estimarem a localização pelo centróide dos RPs. Neste caso, o EM de localização pode ser melhorado pela redução da distância entre RPs, utilizando-se mais RPs no ambiente. No entanto, muitos RPs implicam em mais esforços manuais para criar o conjunto de dados [27].

### B. Erro médio de localização por número de APs e valor do parâmetro k

Foram analisados o efeito do número de APs ($n$) e a influência do número de vizinhos mais próximos ($k$) nas estimativas de localização dos quatro métodos. O método 3-PCA foi testado a partir de três APs, pois esta análise necessita no mínimo de três atributos no conjunto de dados. O EM diminuiu à medida que o número de APs aumentou para os métodos I e PS, mas quando o número de APs é $n≥6$, o EM se torna constante para o método I definido com qualquer $k$, enquanto que o EM de localização para o método PS ainda pôde ser melhorado com $n=8$ (Figura 7).

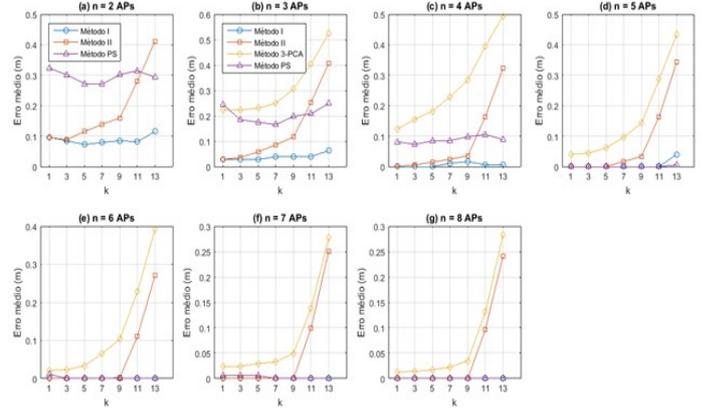

Fig. 7.   Erro médio das estimativas de localização em função do parâmetro $k$ para cada número de pontos de acesso ($n$).

Os métodos II e 3-PCA também tiveram o EM reduzido à medida que o número de APs aumentou, porém houve grande influência dos valores assumidos por $k$, pois à medida que o valor deste parâmetro aumentou os erros tenderam a crescer.

O EM foi nulo para I($n=4$, $k=1$), II($n=4$, $k=1$) (Figura 7c) e PS($n=5$, $k=1$) (Figura 7d). O método 3-PCA($n=8$, $k=1$) apresentou EM de 0,0134 m (Figura 7g). Os métodos I, II e PS também apresentaram EM nulo com outros pares combinados ($n$, $k$), porém os pares apresentados anteriormente possuíram os menores valores possíveis, sendo mais vantajosos. Nota–se que os métodos propostos I e II são equivalentes para qualquer valor de $n$ com $k=1$, pois a média das coordenadas de um único RP são as próprias coordenadas deste RP. Neste caso, o método I é uma simplificação do método II.

Então, o método I($n=4$, $k=1$), com abordagem da análise de quartis na representação dos dados, consiste em uma proposta promissora para localizar objetos em ambientes internos.

### C. Tempo médio de processamento de localização

Para validar um método de localização preciso e com menor tempo de processamento, novos testes foram conduzidos utilizando o método promissor I($n=4$, $k=1$). Foi verificada a precisão alterando o número de leituras de RSSI ($m=20$, $m=15$, $m=10$ e $m=5$) para o cálculo dos quartis dos APs (Figura 8).

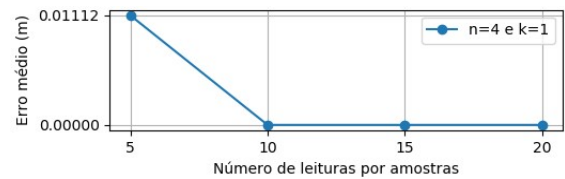

Fig. 8.   Erro de localização com redução do número de leituras de RSSI.

A partir de $m=10$, o EM manteve-se nulo. No entanto, o tempo médio (TM) foi verificado para os testes com $m≥10$. O

TM avaliado é dado em segundos e calculado conforme a Equação 9:

$$TM = \frac{1}{N}\sum_{i=1}^{N} t_i \quad (9)$$

Onde: $t_i$ é o tempo pertinente à criação e classificação de uma instância e $N$ é o número de testes correspondentes.

TABELA II.  TEMPO DE PROCESSAMENTO PARA CADA NÚMERO DE LEITURAS DE RSSI

| Número de leituras (m) | Tempo médio de processamento (s) |
|---|---|
| 10 | 17,27 |
| 15 | 25,58 |
| 20 | 33,76 |

O menor tempo foi de 17,27 segundos, alcançado com $m=10$. Aproximadamente 96% deste tempo envolvem a varredura dos quatro APs e o envio dos dados pela rede Wi-Fi, sendo apenas 0,69 segundo para localizar. Por tanto, o método proposto I($n=4$, $k=1$), com abordagem da análise de quartis na representação dos dados, é o de melhor desempenho e menor tempo de processamento.

## IV. CONCLUSÕES

O método de localização baseado no algoritmo de aprendizado de máquina $k$-Nearest Neighbors ($k$NN) com análise de quartis na representação dos dados, e que verifica as coordenadas do ponto de referência majoritário, possibilitou uma localização de forma precisa em ambientes internos com problemas de instabilidade no RSSI do sinal Wi-Fi.

O desempenho do método proposto supera os métodos relatados na literatura, sendo um sistema de localização de alto desempenho e com rápido tempo de processamento com apenas quatro APs ($n=4$), $k=1$ e $m=10$ leituras por amostra para o cálculo dos quartis. A seleção de apenas uma instância de treinamento ($k=1$) revela que a análise de quartis representa de forma adequada os valores de RSSI.